\documentclass[letterpaper,times]{IONconf}
\usepackage{amsmath}
\usepackage[pdftex]{graphicx}
\usepackage{subcaption}
\usepackage{tabularx,booktabs}
\usepackage{multirow}
\usepackage{indentfirst}
\usepackage{url}
\usepackage{natbib}
\title{A Step Closer Towards  5G mmWave-based Multipath Positioning in Dense Urban Environments}

\author{
    Qamar~Bader, Sharief~Saleh, \textit{Queen's~University, Kingston, ON, Canada}% <- this '%' removes a trailing whitespace
    \vspace{1mm} \\%
    Mohamed~Elhabiby, \textit{Micro~Engineering~Tech~Inc., Calgary, AB, Canada}%
    \vspace{1mm} \\%
    Aboelmagd~Noureldin, \textit{Queen's~University;~Royal~Military~College~of~Canada, Kingston, ON, Canada}%
    }

\begin{document}

\maketitle
% biography section. The * indicates a section excluded from numbering.
\section*{biography}

\biography{Qamar Bader}{Received her B.Sc. degree in electrical engineering from Qatar University, Doha, Qatar, in 2016 and is currently pursuing an MSc degree in electrical engineering at Queen's University, Canada. She is currently a member of the Navigation and Instrumentation Research Lab, RMCC. Her current research interests include 5G mmWave-based positioning and navigation, sensor fusion, deep learning, computer vision, and environment mapping.}

\biography{Sharief Saleh}{Received the B.Sc. and M.Sc. degrees in electrical engineering from Qatar University, Doha, Qatar, in 2016 and 2018 respectively. He is currently pursuing a PhD degree in electrical engineering at Queen's University, Canada. He is currently a member of the Navigation and Instrumentation Research Lab, RMCC. His current research interests include 5G positioning and navigation systems, sensor fusion, and optimal estimators.}

\biography{Mohamed Elhabiby}{Currently an Associate Professor at Ain Shams University, he's also the Executive Vice President and Co-Founder of Micro Engineering Tech Inc. In addition, he's a leader of an archaeological mission in the area of Great Pyramids, Cairo. He has chaired multiple sessions, including the Geocomputations and Cyber Infrastructure Oral Session and the WG 4.1.4: Imaging Techniques. He received the Astech Awards and was named one of the Top 40 under 40 by Avenue Magazine. His expertise is in high precision engineering, instrumentation, mobile mapping, laser scanning, deformation monitoring, and GPS/INS integrations.}

\biography{Aboelmagd Noureldin}{An electrical and computer engineering Ph.D. holder from the University of Calgary, he's a Professor at the Royal Military College of Canada, where he also founded the Navigation and Instrumentation Research Lab. With over 270 published papers, two books, and four book chapters, he's an expert in global navigation satellite systems, wireless location, and navigation, indoor positioning, and multi-sensor fusion. His research focuses on autonomous systems, intelligent transportation, and the vehicular Internet of Things, leading to 13 patents and multiple licensed technologies in the area of position, location, and navigation systems.}

\section*{Abstract}
5G mmWave technology can turn multipath into a friend, as multipath components become highly resolvable in the time and angle domains. Multipath signals have not only been used in the literature to position the user equipment (UE), but also to create a map of the surrounding environment. Yet, many multipath-based methods in the literature share a common assumption, which entails that multipath signals are caused by single-bounce reflections only, which is not usually the case. There are very few methods in the literature that accurately filters out higher-order reflections, which renders the exploitation of multipath signals challenging. This paper proposes an ensemble learning-based model for classifying signal paths based on their order of reflection using 5G channel parameters. The model is trained on a large dataset of $3.6$ million observations obtained from a quasi-real ray-tracing based 5G simulator that utilizes 3D maps of real-world downtown environments. The trained model had a testing accuracy of $99.5\%$. A single-bounce reflection-based positioning method was used to validate the positioning error. The trained model enabled the positioning solution to maintain sub-$30cm$ level accuracy $97\%$ of the time.

\section{INTRODUCTION}
 Accurate positioning and mapping are crucial for the safe and efficient operation of autonomous vehicles, particularly in urban environments. The ability to accurately locate the vehicle within the environment and to understand the layout of the surrounding area is essential for the vehicle to make informed decisions and navigate safely. Urban environments present a number of challenges for autonomous vehicles, such as complex road networks, dynamic traffic conditions, and a high density of obstacles and pedestrians. Without accurate positioning and mapping, autonomous vehicles may struggle to safely navigate these environments and may even pose a risk to other road users \citep{AV}.

\subsection{Problem Statement}
GPS positioning solutions are widely used in autonomous vehicles, but their accuracy and reliability can be affected by the urban environment. In urban areas, high-rise buildings, bridges, and other structures can block or reflect satellite signals, resulting in multipath and shadowing effects. These effects can deteriorate the positioning accuracy or even cause a total signal blockage. Additionally, the crowded radio frequency environment can cause interference with GPS signals \citep{gps}. Inertial Navigation Systems (INS) and perception systems, while useful for autonomous vehicles, have certain limitations when it comes to positioning. INS relies on sensors such as accelerometers and gyroscopes to compute the vehicle's position, velocity, and orientation. However, the accuracy of INS would degrade over time due to the growing IMU biases and the accumulation of errors caused by dead-reckoning. This can make INS less reliable in the long term and it needs to be periodically corrected via an external reliable source \citep{noureldin}. Perception systems, such as cameras and LiDARs, can provide a wealth of information about the vehicle's environment, but they are limited in their ability to provide precise position information. They can be used for localization and mapping, but the accuracy of these systems can be affected by factors such as scene illumination, weather conditions, and the presence of obstacles or occlusions \citep{lidar}.

\subsection{Motivation}
5G mmWave technology is considered a promising alternative as a positioning technology. One of the main features of 5G NR lies in its large bandwidth, which can reach $400$ MHz, enabling accurate time-based measurements such as time of arrival (ToA), time difference of arrival (TDoA), and round-trip time (RTT). Additionally, it enables multipath resolvability in the time domain. 5G also features MIMO capabilities which allow for more accurate estimates of angle-based measurements such as angle of arrival (AoA), and angle of departure (AoD). Massive MIMO capabilities also allow for the ability to resolve multipath signals in the angular domain. Last but not least, the deployment of 5G gNBs is anticipated to occur every $200m$ to $500m$, suggesting an enhanced line-of-sight (LoS) connectivity with the UE. 5G NR has also demonstrated its potential as a mapping technology through its ability to accurately distinguish multipath signals. These signals provide valuable information about the surrounding environment, thereby enabling the creation of environment maps. Multipath signals are useful not only for environment mapping but also for bridging 5G outages that are expected in highly dynamic environments, as well as for providing redundant information to improve the precision of the UE position estimate. However, most of the methods in the literature assume a single-bounce reflection (SBR), rendering multipath signals difficult to use. For instance, the works in \cite{SB1,SB2,SB3,SB4,SB5,SB6} assume SBRs for UE positioning. Additionally, the work in \cite{clock} utilizes SBR in resolving the clock offset between a UE and a BS, enabling precise single-base-station positioning. Therefore, all of the aforementioned methods would directly benefit from methodologies that accurately resolve SBRs from other multipath signals.

\subsection{Contributions}
The ultimate objective of this research is to attain high-precision positioning in dense urban environments where GNSS is denied and vision is degraded. The positioning solution should be robust, continuous, and accurate at the decimeter level. To fulfill this, an order of reflection identification (OoRI) technique is needed to fully exploit multipath signals in multipath-rich environments (i.e. urban canyons).
\noindent Our contributions can be summarised in the following aspects:
\begin{itemize}
  \item Proposal of an ensemble-learning-based order of reflection identifier.
  \item Analysis of positioning errors using a single-bound reflection multipath positioning scheme proposed by \cite{SB0}.
  \item Validation of the proposed approach through a novel quasi-real 5G simulator.
\end{itemize}

\subsection{Significance}
Our proposed method for identifying the order of reflection for multipath signals has the potential to revolutionize the field of positioning and mapping. The ability to accurately identify the order of reflection can enable realizing many positioning and mapping methods in the literature. Furthermore, the proposed method can also be used to improve the accuracy of mapping systems that rely on wireless signals, by providing redundant information about the location and movement of objects.

\subsection{Related Work}
There are limited works in the literature that addresses identifying the order of reflection of multipath signals. The authors in \citep{RW1} propose a two-step proximity detection scheme to detect and discard multiple-bound scattering paths in order to identify the order of reflection. First, they find the centroid of the line of possible mobile device location (LPMD) using a normalized path weighting factor. Second, they calculate the normalized euclidean distance between the midpoint of each path and the estimated centroid, paths whose distance exceeds a pre-determined threshold being considered as multiple-bound scattering paths. The proposed approach is computationally intensive, as it necessitates the calculation of the LPMD for all paths. This may pose challenges in the real-time implementation of the method. Additionally, the weighting strategy employed in the approach, which is based on the assumption that multiple-bound scattering paths have a larger TOA compared to one-bound scattering paths, may not always hold true in all scenarios. In \citep{RW2}, authors propose a method based on the observation that diffuse scattering and multiple reflection channels have significantly lower RSS than LOS and single-bounce specular reflection paths. The strongest components are chosen from the received paths using a predetermined threshold in order to filter out these interfering paths. However, this approach is unnecessarily discarding useful information. Specifically, it fails to take into account that single-bounce diffracted signals can be useful for positioning and mapping applications \citep{SB0}. Research in the field has demonstrated the utility of both types of signal interactions for these purposes, thus, relying solely on single-bounce specular reflections may not provide a comprehensive representation of the environment.

\subsection{Paper Organization}
The remainder of the paper is organized as follows: Section II establishes the system model and the foundations of 5G positioning. Section III exhibits the proposed order of reflection identification scheme. Section IV provides details on the experimental setup and presents the results along with discussions. Finally, Section VI concludes the paper.

\section{System Model}
\subsection{5G System Model}
To enable more accurate positioning measurements, 5G NR features dedicated reference signals \citep{refsignals}. These signals are known as the UpLink Sounding Reference Signal (UL-SRS) and the DownLink Positioning Reference Signal (DL-PRS). Through the correlation of the received reference signal and the original reference signal configuration, such signals assist in the estimation of ToA, TDoA, and RTT. Typically, UL-AoA is estimated using UL-SRS in conjunction with multiple signal classification (MUSIC) or estimation of signal parameters using rotational invariance techniques (ESPRIT) \citep{music}. On the other hand, DL-AoD estimation often occurs during the beamforming (BF) training sequence. This paper uses DL-PRS signals to compute AoD and RTT since the estimated position is computed at the UE side rather than the network. The position of the UE in relation to the location of the base station (gNB) in a 3D Cartesian coordinate system is mathematically represented by the following equations:

\begin{equation}
        \boldsymbol{p}'=\boldsymbol{p}_b' + d'
       \begin{pmatrix}
         \sin\alpha \cos\phi\\
         \cos\alpha \cos\phi\\
         \sin(\phi)
        \end{pmatrix} 
\end{equation}

\noindent Where $\boldsymbol{p}' = [x,y,z]^T$ and $\boldsymbol{p}_b' = [x_b,y_b,z_b]^T$ represent the 3D Cartesian coordinates of the UE and the BS deceptively, $d'$, $\alpha$, and $\phi$ represent the spherical coordinates of the UE, with $d'$ being the 3D distance between the UE and the BS. The angle between the UE and the BS is denoted by $\alpha$, while the elevation angle is denoted by $\phi$. If the vehicle's height is known, the 2D position of the vehicle can be calculated as seen in (\ref{Hybrid}).

\begin{equation}
\label{Hybrid}
    \begin{split}
        &\boldsymbol{p} = \boldsymbol{p_b} + d  
        \begin{pmatrix}
         \sin\alpha \\
         \cos\alpha
        \end{pmatrix} \\
        &d=\sqrt{r^2-\Delta z^2}
    \end{split}
\end{equation}

\noindent Where $\boldsymbol{p}= [x,y]^T$ denote the 2D positioning of the UE; $d$ is the 2D range between the UE and the gNB, and $\Delta z$ is the height difference between the UE and the gNB.

\subsection{Multipath Positioning} \label{MPP}

\begin{figure}[ht]
	\centering
	\includegraphics[width=4in]{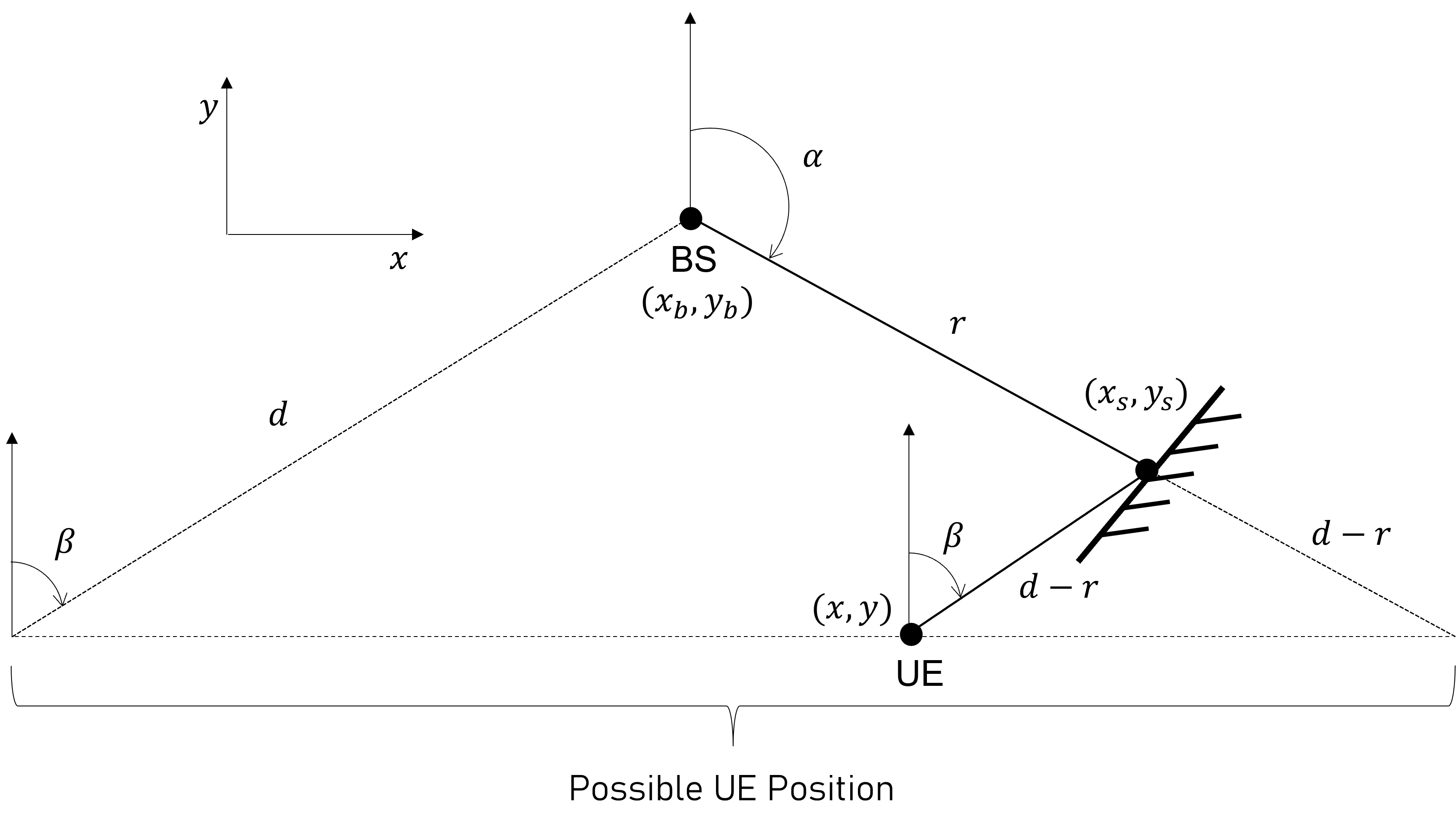}
	\DeclareGraphicsExtensions.
	\caption{System model of a single-bounce reflection scenario.}
	\label{fig:MPP}
\end{figure}

A novel algorithm for multipath positioning is proposed in \citep{SB0}. The algorithm estimates the possible region of the UE position by using the AOD, denoted as $\alpha$, AOA $\beta$, and the distance $d$, of the strongest propagation path, as shown in Fig.\ref{fig:MPP}.

\noindent The coordinates of the scatterer, $ \boldsymbol{p_{s}}=[x_s,y_s]^T$, and the UE, $\boldsymbol{p}$, are given by the following equations:
\begin{equation} 
\boldsymbol{p_{s}} = \boldsymbol{p_{b}} + r 
\begin{pmatrix}
\sin\beta \\
\cos\beta
\end{pmatrix}, \qquad r\in(0,d)
\end{equation}

\begin{equation} 
\boldsymbol{p} = \boldsymbol{p_{s}} - (d - r)
\begin{pmatrix}
\sin\alpha \\
\cos\alpha
\end{pmatrix} , \qquad r\in(0,d)
\end{equation}

\noindent Where $r$ is the distance between the BS and the scatterer, and $d$ is the distance of the strongest path. The possible position of the UE can be described by the straight-line equation:
\begin{equation} 
y=k(\alpha,\beta)x+b(\alpha,\beta,d),
\end{equation}
 where,
 \begin{equation} 
k(\alpha,\beta) = \frac{\cos \alpha + \cos \beta}{\sin \alpha + \sin \beta},
\end{equation}
and,
\begin{equation} 
b(\alpha,\beta,d) = -k(\alpha,\beta)(x_{b} - d\sin\alpha) + y_{b} - d\cos\alpha.
\end{equation}

\noindent This implies that if there is knowledge about two propagation paths from the UE, the position of the UE can be estimated as the intersection of two lines.

\section{Order of Reflection Identification}
\begin{figure}[h]
	\centering
	\includegraphics[width=3in]{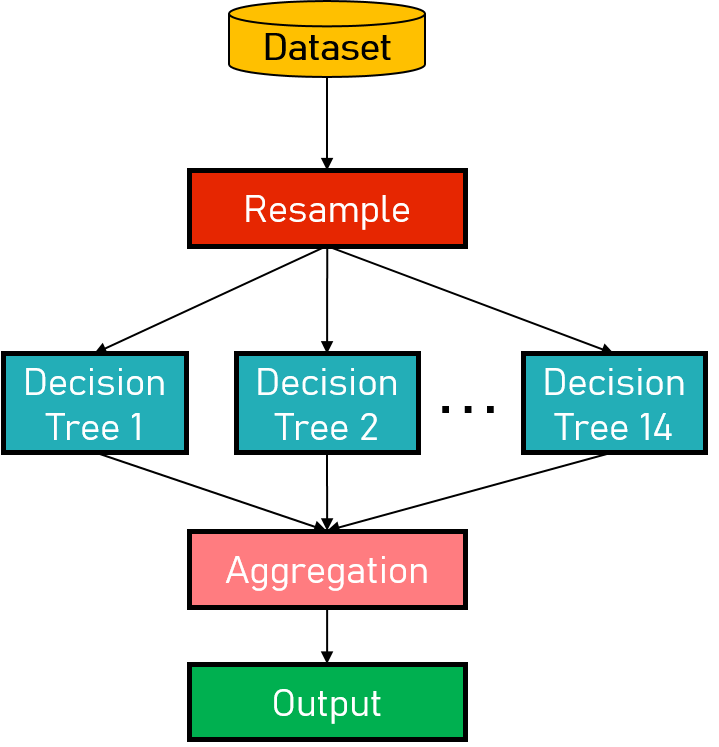}
	\DeclareGraphicsExtensions.
	\caption{Block diagram of the ensemble learning model.}
	\label{ensemble}
\end{figure}

\begin{figure}[h]
	\centering
	\includegraphics[width=3.5in]{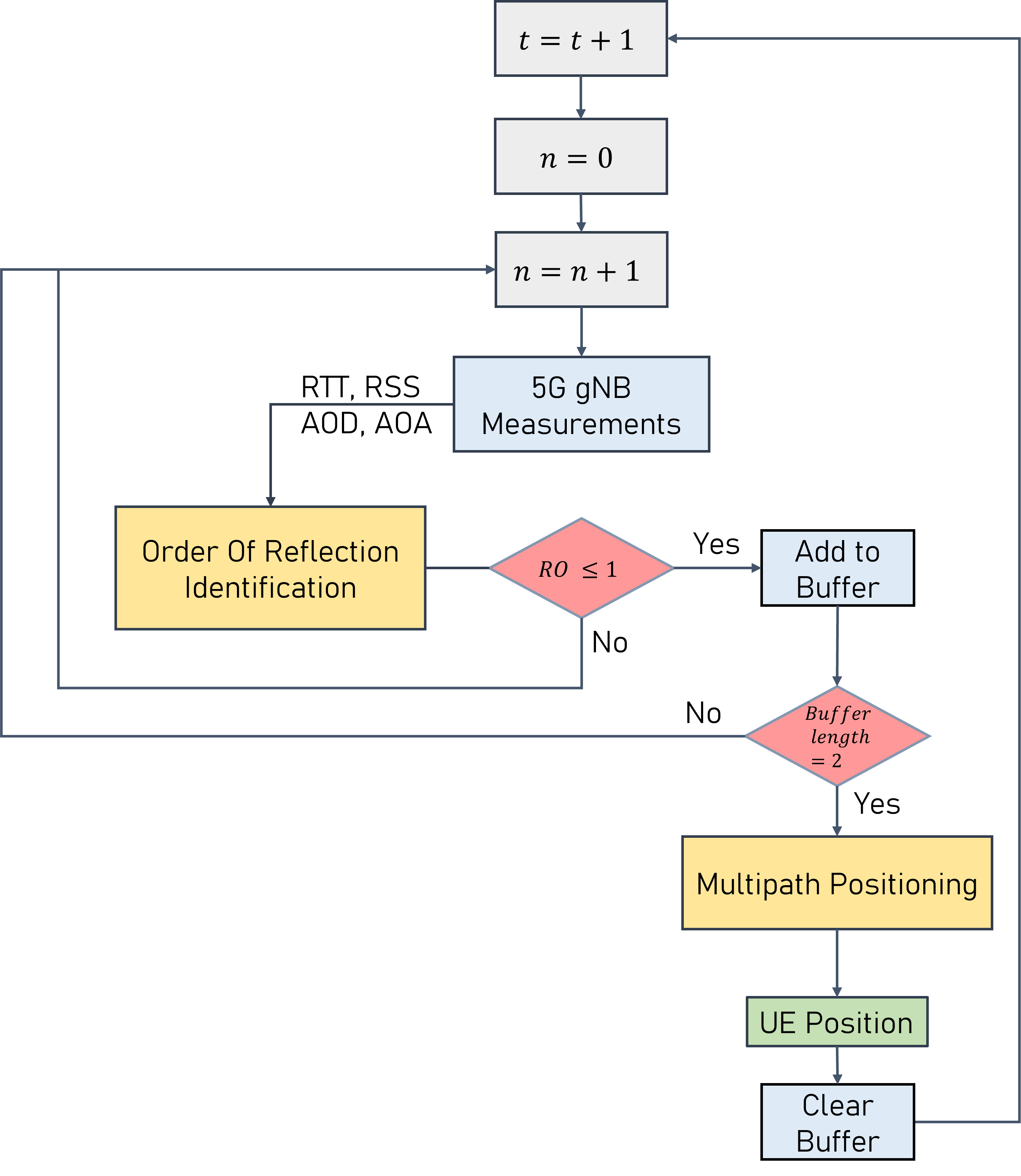}
	\DeclareGraphicsExtensions.
	\caption{Block diagram of the proposed positioning solution.}
	\label{method}
\end{figure}

Identifying the order of reflection of multi-path signals is not a trivial problem. This is due to the fact that such signals undergo various interactions with the environment such as reflection, diffraction, and refraction. A single time-based, angle-based, or power-based observation might correspond to various combinations of these interactions. To address such a complex classification task a simple learning-based approach is not efficient, therefore, ensemble algorithms are investigated. Ensemble learning is a machine learning technique that combines several base models in order to produce a more accurate, robust, and reliable model \citep{Ensemble}. It is used to improve the performance of a model, assign a confidence score to the decision made by the model, and data fusion. The proposed model is summarised in Fig.\ref{ensemble}. Bagging, which stands for bootstrap aggregating, is the ensemble method used in this work. It is a technique used to reduce the variance of a base model by averaging its predictions across multiple samples of data. Bagging works by randomly selecting subsets of data with replacement and training individual models on each subset. Once the models are trained, a voting procedure takes place to consolidate their predictions. In this work, the ensemble model was constructed by aggregating the predictions of fourteen decision tree models \citep{DecisionTree}. The dataset under examination consists of around $3.6M$ observations, with features including TOA, AOA, AOD, and RSS. The ground truth labels are provided by a quasi-real 5G simulator as described in detail in section \ref{Exp}. The labels comprise seven classes representing the reflection order (RO) of each multipath signal, with class 0 denoting an LoS signal and class 7 denoting a six-bounce NLoS signal. The data was shuffled before being split into training, validation, and testing sets, with proportions of $60\%$, $20\%$, and $20\%$ respectively. This allows for an accurate evaluation of models developed using the dataset, as the models will be tested on a diverse set of unseen data, ensuring its generalizability. The large size of the dataset also allows for a high level of statistical power, which is important for uncovering meaningful patterns and relationships within the data. For validation, 5-fold cross-validation is utilized to detect overfitting and evaluate the generalization performance of the model \citep{crossV}. In order to verify the premise of the paper, the single bounce-based multipath positioning solution proposed by \cite{SB0} is implemented using data from the proposed method. The general framework is summarized in Fig.\ref{method}. For every time instant $t$ there exist many signal reflections. The 5G measurements are fed into the OoRI method, which filters out reflections with ROs greater than one (i.e. multiple-bounce reflections). As a result, once two reflections satisfy the foregoing condition, the corresponding 5G measurements are fed into the multipath positioning method for UE position computation. If no reflection satisfies the first condition, the positioning algorithm uses the strongest two paths, which may result in errors in the final position solution.

\section{Experimental Road Tests and Results}
\subsection{Experimental Road Tests} \label{Exp}
\begin{figure}
\centering
\begin{subfigure}{.5\textwidth}
  \centering
  \includegraphics[width=.98\linewidth]{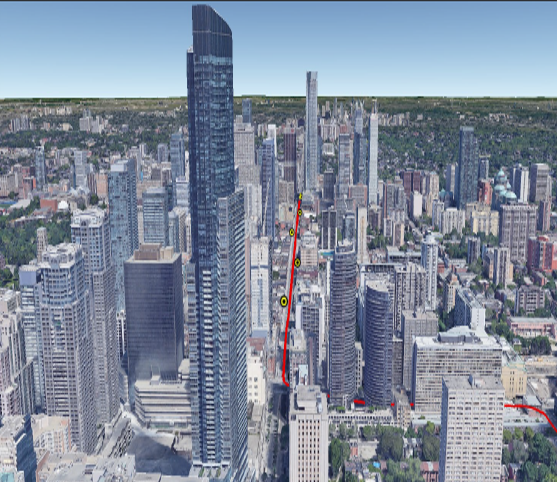}
  \caption{Google Earth}
  \label{GoogleEarth}
\end{subfigure}%
\begin{subfigure}{.5\textwidth}
  \centering
  \includegraphics[width=.98\linewidth]{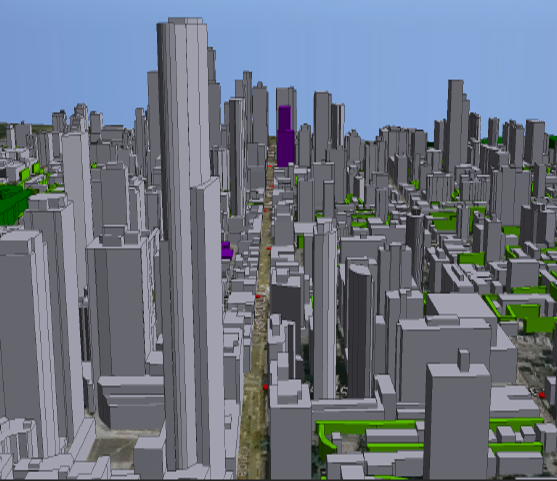}
  \caption{Siradel simulation environment}
  \label{Siradel}
\end{subfigure}
\caption{A snapshot of a segment from downtown Toronto, ON, on (a) Google Earth, vs. (b) Siradel simulation environment}
\label{GE Vs Si}
\end{figure}

Siradel, a 5G simulation environment that mimics 5G measurements based on ray-tracing capabilities, was used to test the proposed method. The simulation environment is made up of 3D maps that resemble real-world settings and were obtained from LiDAR scans of downtown Toronto. Fig.\ref{GoogleEarth} depicts a Google Earth snapshot of a segment from downtown Toronto. A snapshot of the corresponding area in the Siradel environment is depicted in Fig.\ref{Siradel}. This suggests that acquired multipath signals are likely to approximate realistic signal propagation, as utilizing a map of the real downtown environment provides accurate information on the physical properties of the environment. For instance, building heights, street widths, and material compositions, which affect the propagation of 5G signals. 

\begin{figure}[hb]
\centering
\includegraphics[width=6.2in]{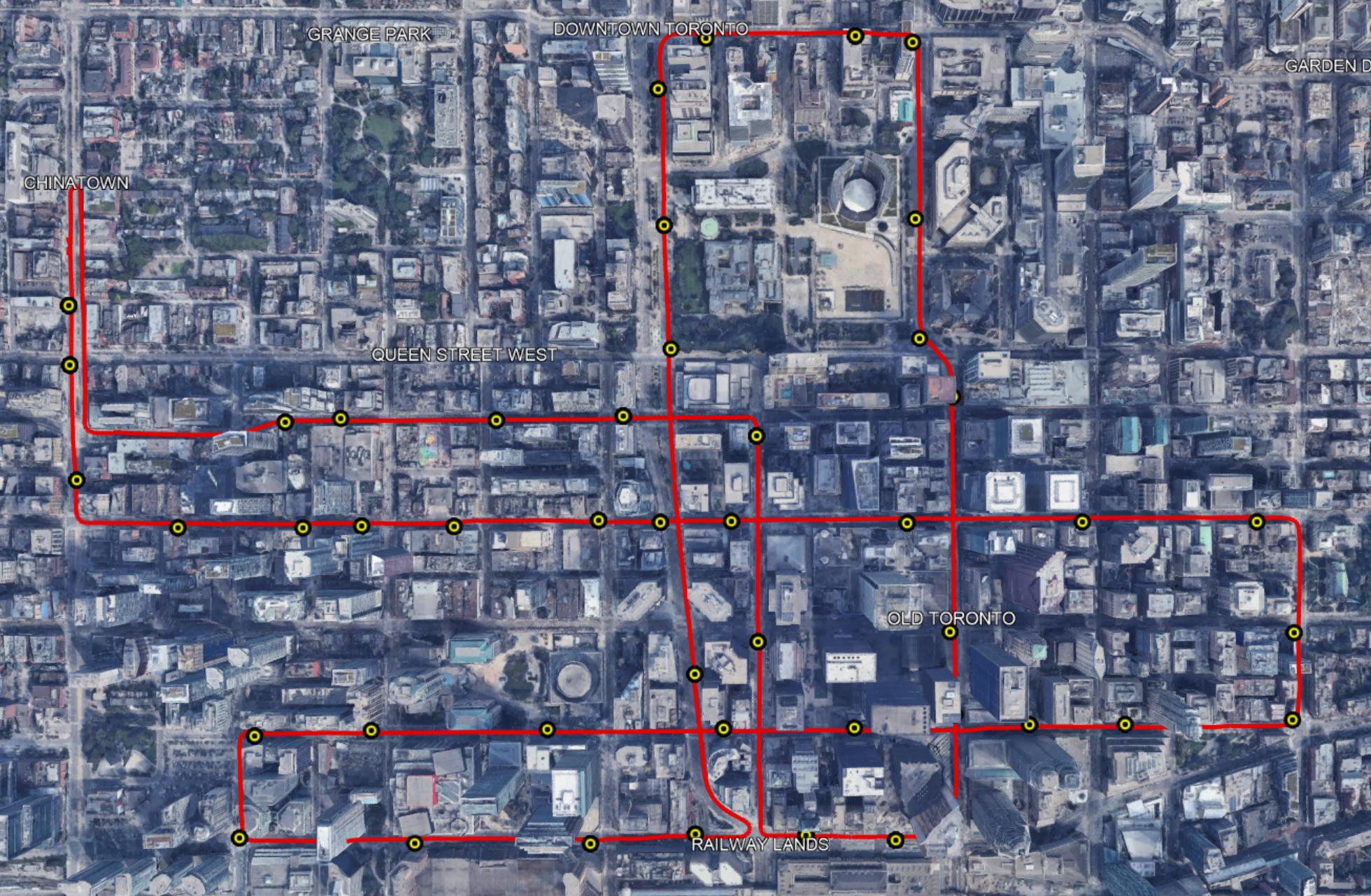}
\caption{Downtown Toronto Trajectory (Red), and 5G gNBs (Yellow circles).}
\label{Traj}
\end{figure}

A reference trajectory of the UE and corresponding positions of the connected gNBs was imported to Siradel to acquire the 5G measurements. A NovAtel tactical grade IMU/GNSS positioning solution was placed on a vehicle to collect the trajectory, which was driven over a distance of $9km$ inside Toronto's downtown region for approximately $1$ hour and $13$ minutes as seen in Fig.\ref{Traj}. Additionally, gNBs were placed along the trajectory, $250m$ apart, and $4m$ lateral distance from the UE's trajectory of motion. 
Lastly, the carrier frequency was set at $28GHz$ with a $400MHz$ bandwidth to obtain mmWave transmissions. The BS was equipped with an $8x1$ ULA while the UE had access to an omnidirectional antenna.

% \begin{figure}[ht]
% 	\centering
% 	\includegraphics[width=4in]{Figures/GE vs Siradel.pdf}
% 	%\DeclareGraphicsExtensions.
% 	\caption{Downtown Toronto, ON, Google Earth (Top) vs Siradel simulation tool (Bottom).}
% 	\label{GoogleEarth vs Siradel}
% \end{figure}

\subsection{Results and Discussions}
The proposed method was trained using the model described in Fig.\ref{ensemble} using Matlab's machine learning toolbox. To evaluate the trained model, first, we present the classification error, and then  evaluate its positioning performance utilizing the multipath positioning method by \cite{SB0} within the framework described in Fig.\ref{method}.

\begin{figure}[ht]
\centering
\includegraphics[width=4in]{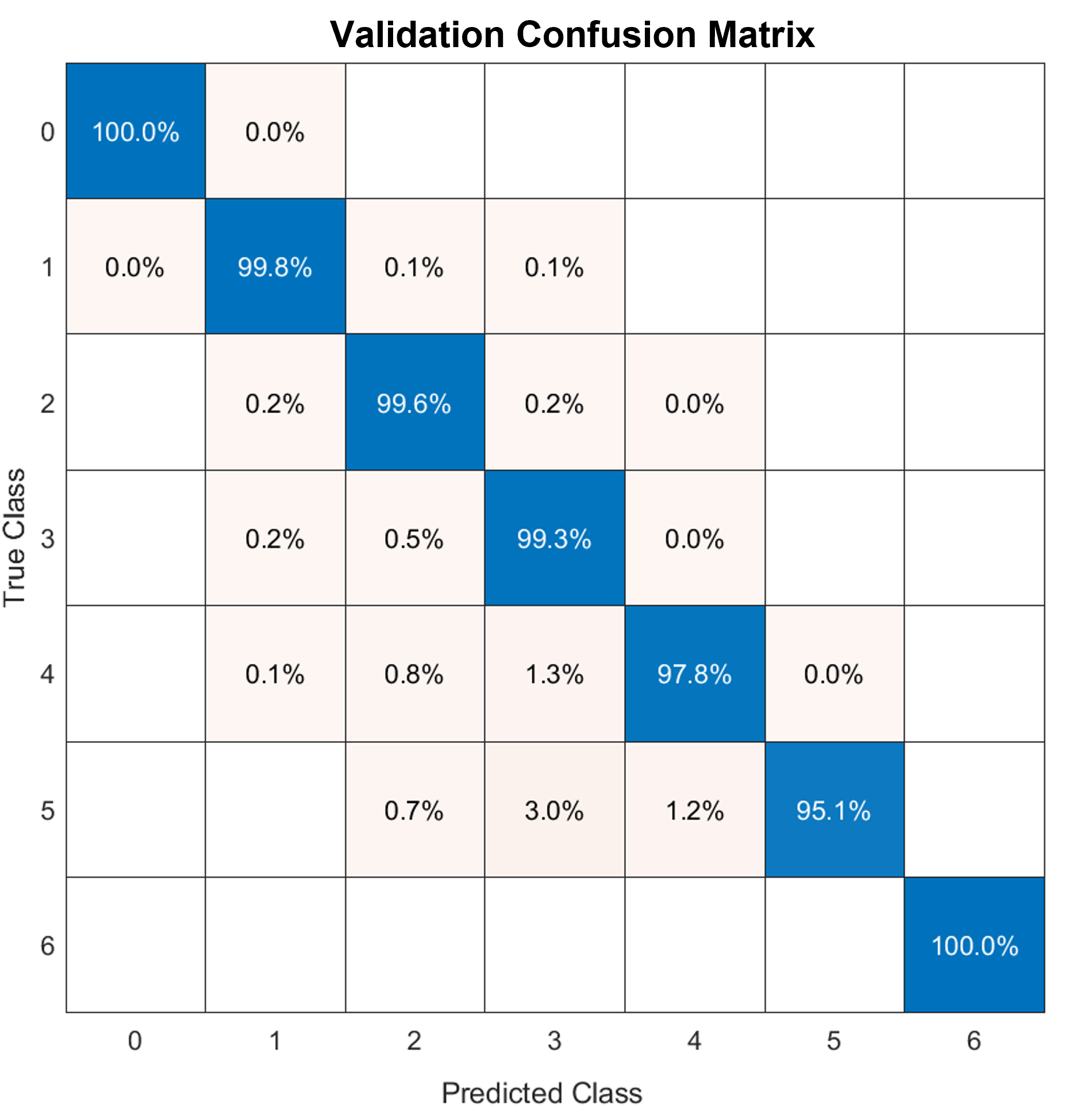}
\caption{Validation confusion matrix of the trained model.}
\label{confusion}
\end{figure}

Fig.\ref{confusion} exhibits the validation confusion matrix where the diagonal elements of the table represent the number of correct classifications, while the off-diagonal elements represent the number of incorrect classifications.  False positives are reported by the off-diagonal horizontal elements of the confusion matrix, while false negatives are reported vertically. The percentage of false positives exhibits the percentage of SBR wrongly classified as higher-order reflections, which induces unnecessary 5G outages. On the other hand, the percentage of false negatives represents the percentage of higher-order reflections being wrongly classified as SBRs, which will cause positioning errors. Overall, the trained model was able to achieve a validation accuracy of $99.5\%$, and testing accuracy of $99.6\%$ resulting in a highly accurate model. Additionally, the close proximity of the two accuracy scores suggests that the model is not overfitting to the training data. 

\begin{figure}[ht]
\centering
\includegraphics[width=6.5in]{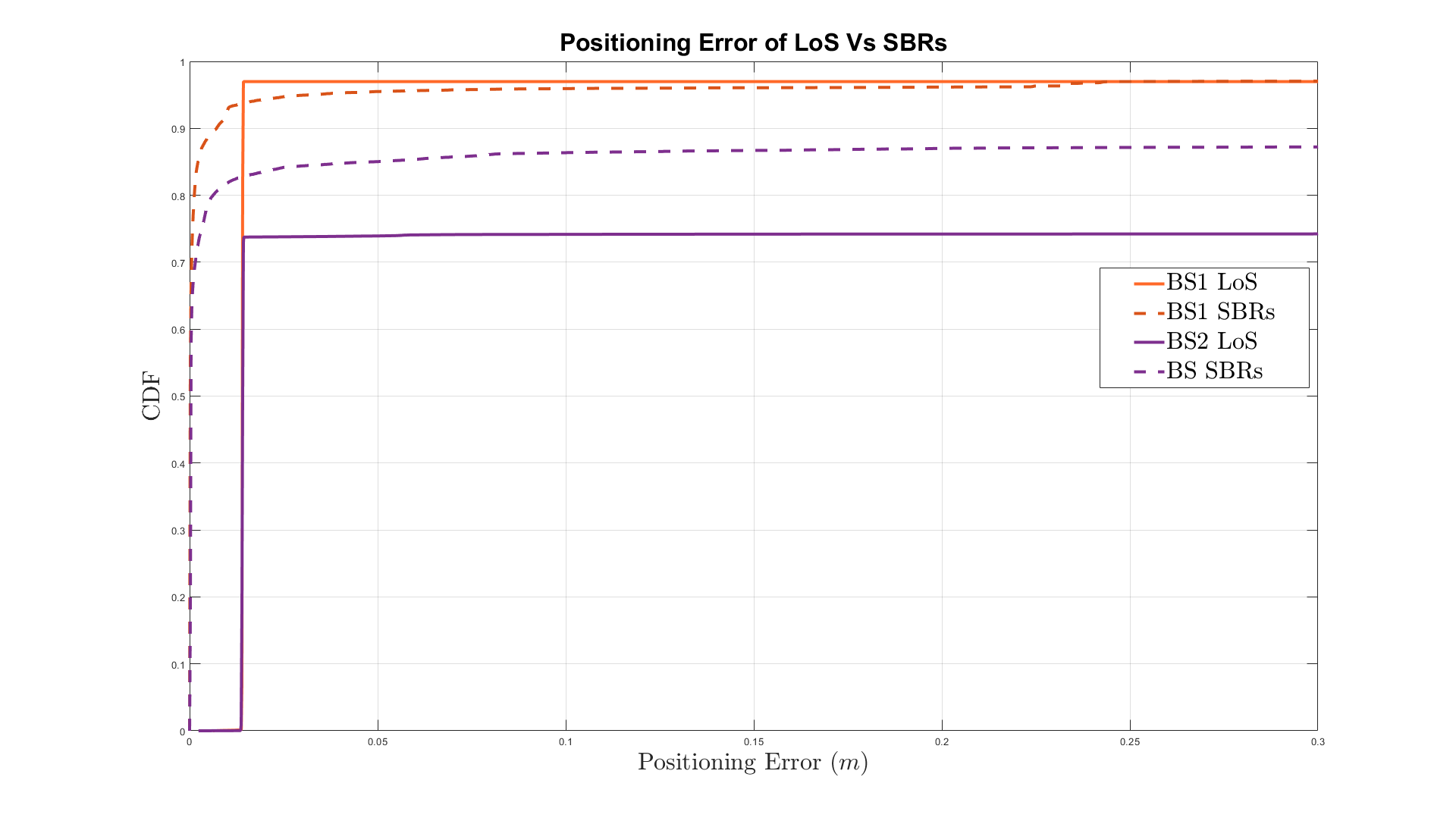}
\caption{CDF of the 2D positioning errors of LoS Vs. SBR-based positioning.}
\label{cdf}
\end{figure}

\begin{figure}[ht]
\centering
\includegraphics[width=6.5in]{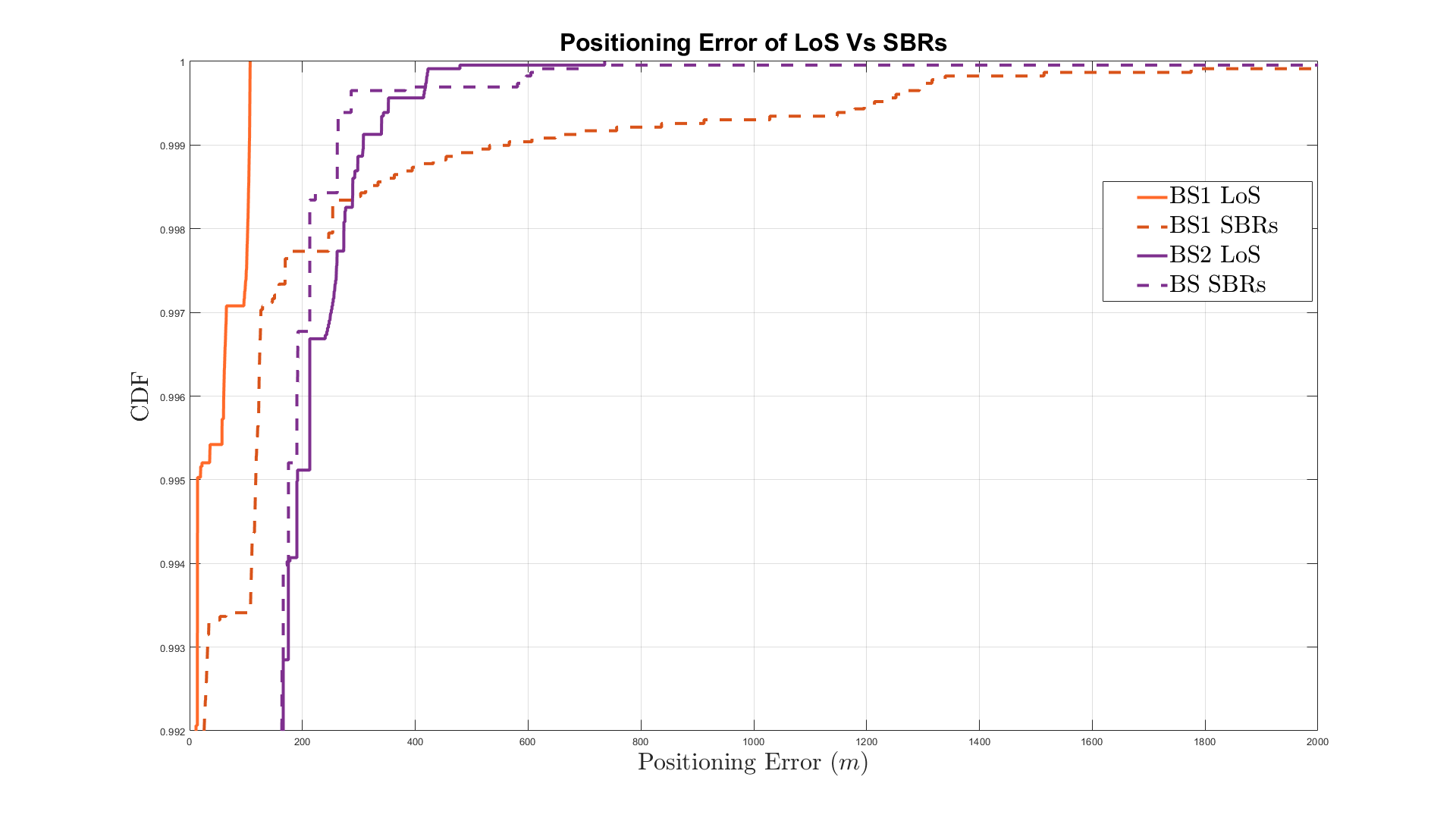}
\caption{Zoomed CDF of the 2D positioning errors of LoS Vs. SBR-based positioning.}
\label{cdf2}
\end{figure}

The positioning performance of the LoS and NLoS 5G measurements are evaluated by plotting the cumulative distribution function (CDF) of the positioning error as seen in Fig.\ref{cdf}. The positioning error statistics for both 5G-based hybrid positioning using \ref{Hybrid} and SBR-based positioning, as described in section \ref{MPP}, are summarized in Table \ref{results_table}. The presented results show the variation between the closest gNB denoted gNB1, which has a high probability of LoS connectivity, and the second closest gNB denoted gNB2, which has a lower chance of LoS connectivity. Overall, it is clear that SBR-based positioning has almost identical error statistics to LoS-based positioning while employing gNB1 measurements, which can maintain sub-$30cm$ accuracy for $97\%$ of the time. However, it is evident that the maximum error of positioning based on SBR is significant. This is because SBRs are not always available, even in dense urban areas, forcing the method to rely on higher-order reflection, resulting in incorrect UE position computation. When analyzing the positioning error of the second closest gNB, SBR-based positioning exhibits superior results as it can sustain sub-$30cm$ error for $87\%$ of the time compared to merely $74\%$ for the positioning scheme utilizing LoS signals. This is because LoS signals become less likely as we move away from the base station, yet, multipath reflections remain available. Fig. \ref{cdf2} shows a closer look at the high-error end of the CDF. It can be seen that while SBR positioning outperforms LoS positioning at low error ranges, yet, their errors during outages are much higher compared to LoS-based positioning. Proving the fact that relying solely on higher-order reflections for positioning could cause serious positioning errors. Hence, the fusion with LoS measurements is crucial. 

\begin{table}[ht]
	\caption{2D Positioning Error Statistics With 5G hybrid 
                 Positioning Vs SBR-based positioning.}
	\label{results_table}
        \centering
	\begin{tabularx}{0.7\columnwidth}{@{}l*{5}{c}c@{}}
		\toprule
		&Error      &\multicolumn{2}{c}{gNB1}     &\multicolumn{2}{c}{\hspace{20pt}gNB2}\\    
        &\hspace{20pt}Type\hspace{20pt}         &\hspace{20pt}LoS\hspace{20pt}         &\hspace{20pt}SBRs\hspace{20pt}           &\hspace{20pt}LoS\hspace{20pt}     &SBRs\\
		\midrule
		&RMS            & $6.3m$     & $51m$         &$45m$   &$47m$\\ 
		&Max            & $107m$     & $4734m$       &$735m$  &$3124m$\\ 
		&sub $2 m$      & $97\%$     & $99\%$        &$82\%$  &$87\%$\\ 
		&sub $1 m$      & $97\%$     & $99\%$        &$81\%$  &$87\%$\\ 
		&sub $30 cm$    & $96.9\%$   & $98.8\%$        &$74\%$  &$87\%$ \\
		\bottomrule
	\end{tabularx}
\end{table}

\begin{figure}[ht]
\centering
\includegraphics[width=7in]{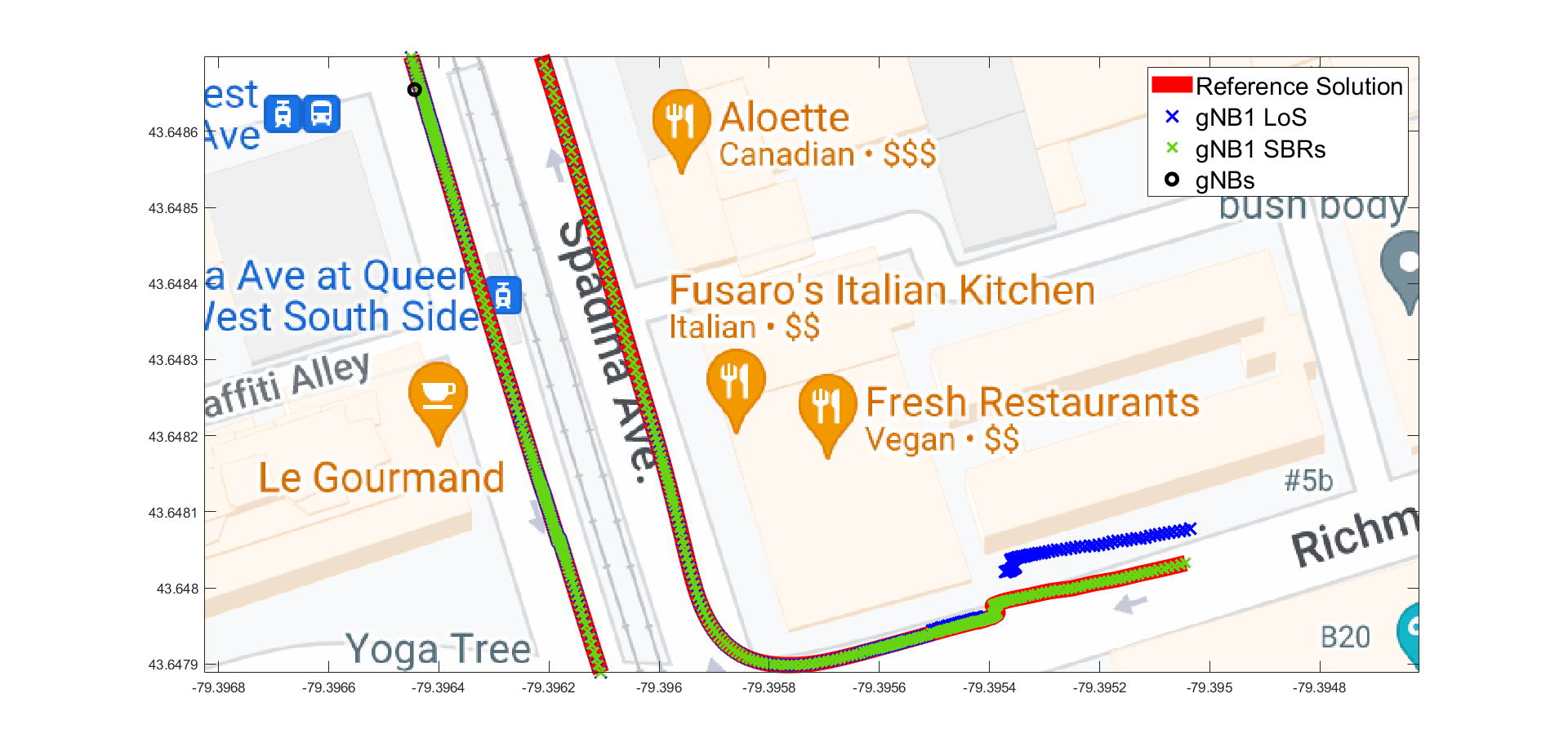}
\caption{Close-up scenario that showcases the capability of multipath positioning accuracy during 5G outage while the UE is connected to the closest BS (gNB1).}
\label{gNB1-CU1}
\end{figure}

\begin{figure}[ht]
\centering
\includegraphics[width=7in]{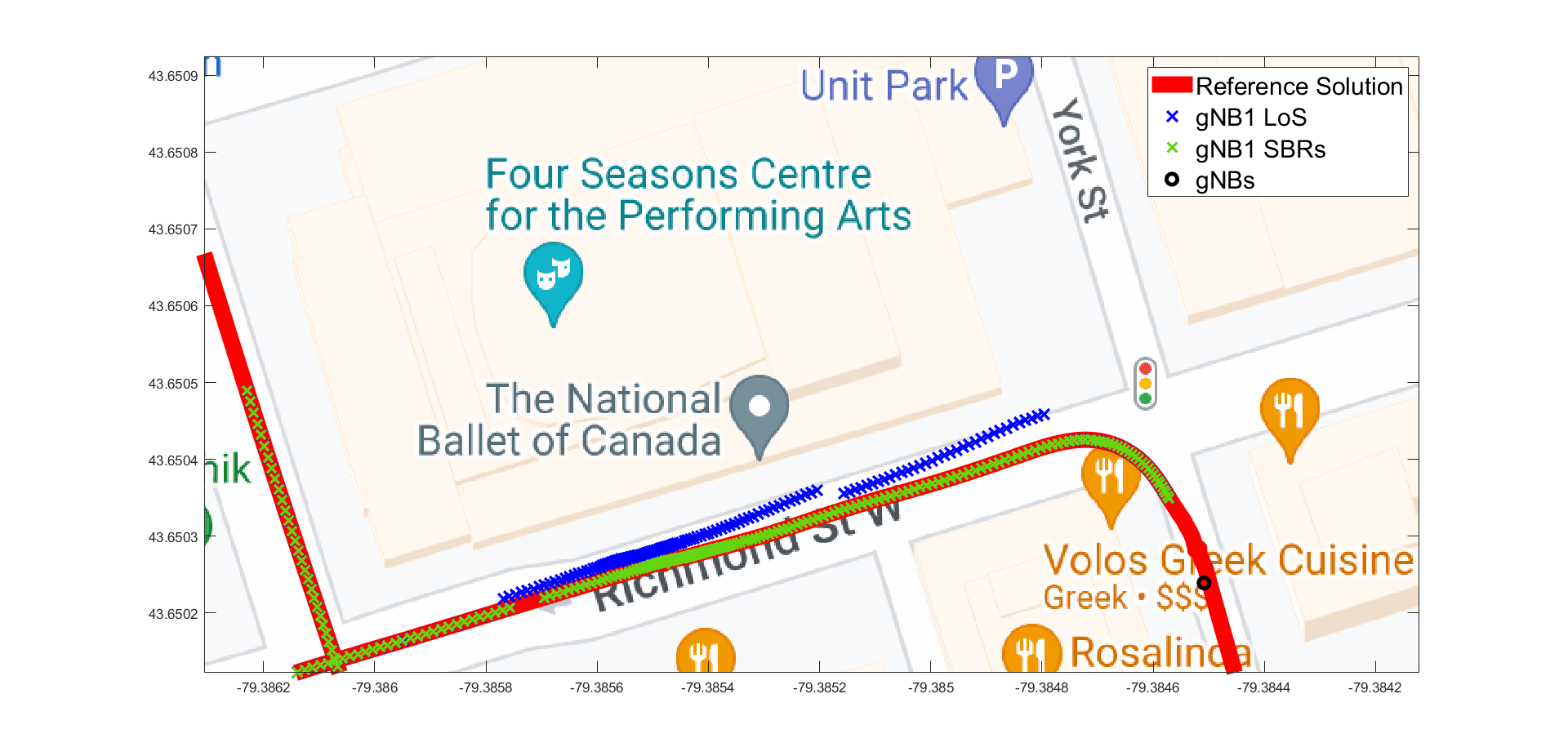}
\caption{Close-up scenario that showcases the capability of multipath positioning accuracy during 5G outage while the UE is connected to the second closest BS (gNB2).}
\label{gNB2-CU1}
\end{figure}

A Close-up sample of the positioning solution of LoS and SBR-based positioning solutions for both gNB1 and gNB2 are shown in Figs.\ref{gNB1-CU1} and \ref{gNB2-CU1} respectively. In the event of 5G outages, multipath signals can bridge the gap while maintaining high positioning accuracy. This demonstrates the availability of SBR signals over LoS in dense urban areas.

\begin{figure}[h]
\centering
\includegraphics[width=7in]{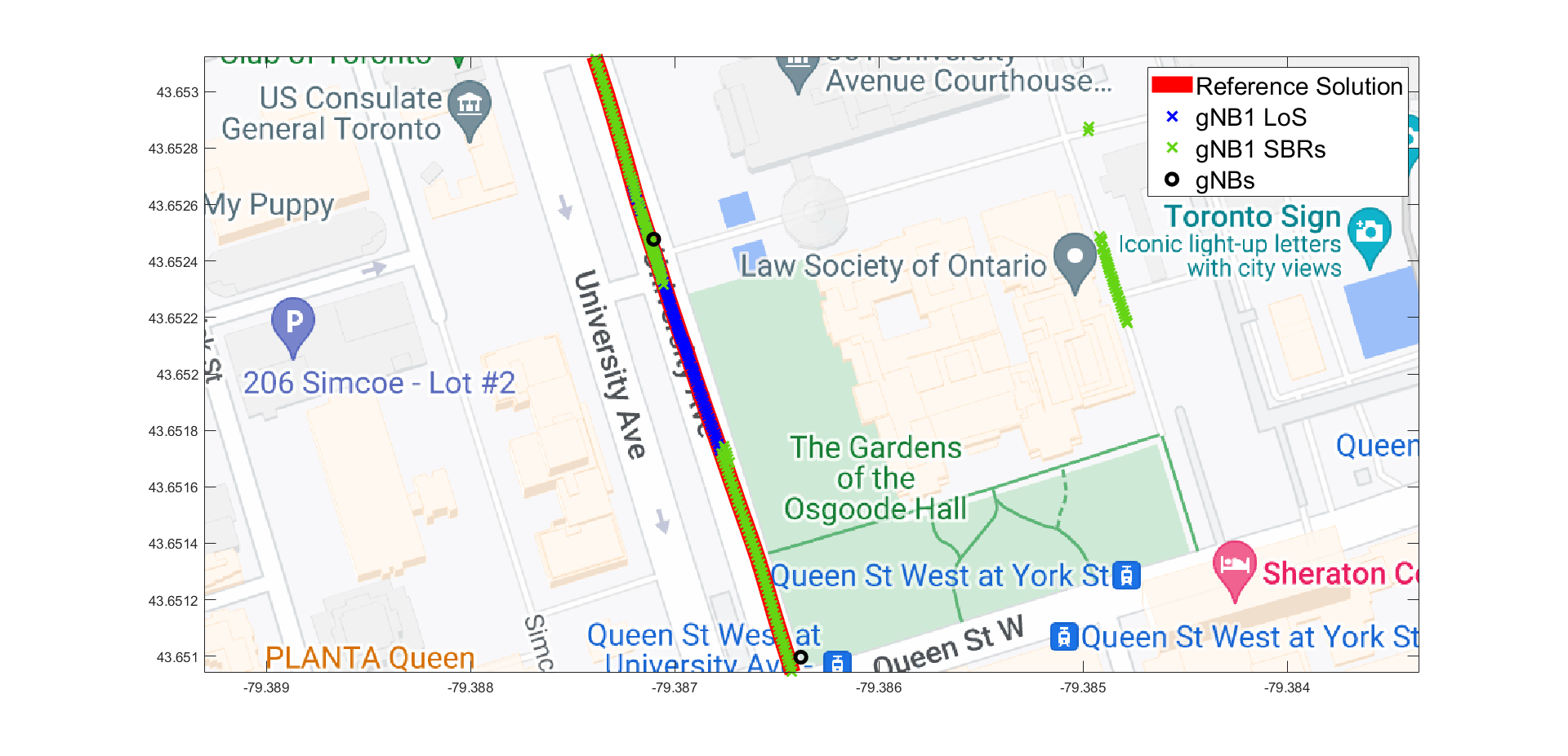}
\caption{A close-up scenario in which LoS measurements are available but sufficient SBRs are not.}
\label{los}
\end{figure}

In rare cases, the number of SBRs was not efficient as to enable multipath positioning, causing erroneous positioning solutions as seen in Fig. \ref{los}. In the meantime, LoS communication with the UE exists which demonstrates the significance of fusing LoS and SBRs for more accurate, highly-precise position estimates. SBR-based positioning is obviously intended to supplement LoS-based positioning rather than replace it. They are, however, investigated separately to demonstrate the potential of using multipath signals for positioning.

\section{Conclusion}
In conclusion, we presented a novel method for accurately classifying multipath signals based on their order of reflection (i.e. LoS, single-bounce, double-bounce, etc.) using 5G channel parameters such as ToA, AoA, AoD, and RSS. Our proposed model was trained on a dataset of $3.6$ million observations obtained from a quasi-real 5G simulator using an ensemble learning technique with bagging of $14$ decision tree models. The model demonstrated its reliability as a RO classifier by achieving validation accuracy of $99.5\%$ and testing accuracy of $99.6\%$. A multipath positioning technique that employs channel parameters of two SBRs to determine the position of the UE was also used to demonstrate the effectiveness of the proposed classifier in a positioning setting. Whilst using the proposed OoRI method, the positioning error was maintained below $30$ cm for $98.8\%$ of the time, which was an improvement over LOS positioning. To further demonstrate the trained model's capability, we investigated its positioning error while using measurements from the second closest gNB, which has a lower chance of LoS connectivity. The SBR-based positioning was observed to maintain sub-30 cm accuracy for $87\%$ of the time, which significantly outperformed LOS positioning. The findings of this research have the potential to vastly improve existing positioning and mapping methods that rely solely on SBRs. Overall, this study lays a solid foundation for future research in multipath signal classification and positioning.

\clearpage
\section*{ACKNOWLEDGEMENTS}
This research is supported by grants from the Natural Sciences and Engineering Research Council of Canada (NSERC) under grant numbers: RGPIN-2020-03900 and ALLRP-560898-20. 

\bibliographystyle{apalike}
\bibliography{References}

\begin{thebibliography}{}

\bibitem[Benedek et~al., 2021]{lidar}
Benedek, C., Majdik, A., Nagy, B., Rozsa, Z., and Sziranyi, T. (2021).
\newblock Positioning and perception in {LiDAR} point clouds.
\newblock {\em Digital Signal Processing}, 119:103193.

\bibitem[Jing et~al., 2022]{gps}
Jing, H., Gao, Y., Shahbeigi, S., and Dianati, M. (2022).
\newblock Integrity monitoring of {GNSS/INS} based positioning systems for
  autonomous vehicles: State-of-the-art and open challenges.
\newblock {\em IEEE Transactions on Intelligent Transportation Systems}.

\bibitem[Kakkavas et~al., 2020]{SB4}
Kakkavas, A., Garc{\'{\i}}a, M. H.~C., Seco{-}Granados, G., Wymeersch, H.,
  Stirling{-}Gallacher, R.~A., and Nossek, J.~A. (2020).
\newblock Position information from single-bounce reflections.
\newblock {\em CoRR}, abs/2012.01597.

\bibitem[Kakkavas et~al., 2021]{SB5}
Kakkavas, A., Garc{\'\i}a, M. H.~C., Seco-Granados, G., Wymeersch, H.,
  Stirling-Gallacher, R.~A., and Nossek, J.~A. (2021).
\newblock Position information from reflecting surfaces.
\newblock {\em IEEE Wireless Communications Letters}, 10(6):1300--1304.

\bibitem[Kakkavas et~al., 2019]{clock}
Kakkavas, A., Garc{\'\i}a, M. H.~C., Stirling-Gallacher, R.~A., and Nossek,
  J.~A. (2019).
\newblock Performance limits of single-anchor millimeter-wave positioning.
\newblock {\em IEEE Transactions on Wireless Communications},
  18(11):5196--5210.

\bibitem[Kulmer et~al., 2017]{SB6}
Kulmer, J., Hinteregger, S., Gro{\ss}windhager, B., Rath, M., Bakr, M.~S.,
  Leitinger, E., and Witrisal, K. (2017).
\newblock Using decawave {UWB} transceivers for high-accuracy
  multipath-assisted indoor positioning.
\newblock In {\em 2017 IEEE International Conference on Communications
  Workshops (ICC Workshops)}, pages 1239--1245. IEEE.

\bibitem[Li and Wang, 2021]{SB3}
Li, B. and Wang, X. (2021).
\newblock Rigid body localization and environment sensing with {5G} millimeter
  wave mimo.
\newblock In {\em 2021 IEEE 94th Vehicular Technology Conference
  (VTC2021-Fall)}, pages 1--5.

\bibitem[Lin et~al., 2018]{RW2}
Lin, Z., Lv, T., and Mathiopoulos, P.~T. (2018).
\newblock {3-D} indoor positioning for millimeter-wave massive mimo systems.
\newblock {\em IEEE Transactions on Communications}, 66(6):2472--2486.

\bibitem[Miao et~al., 2007]{SB0}
Miao, H., Yu, K., and Juntti, M.~J. (2007).
\newblock Positioning for {NLOS} propagation: Algorithm derivations and
  cramer--rao bounds.
\newblock {\em IEEE Transactions on Vehicular Technology}, 56(5):2568--2580.

\bibitem[Mogyorósi et~al., 2022]{refsignals}
Mogyorósi, F., Revisnyei, P., Pašić, A., Papp, Z., Törös, I., Varga, P.,
  and Pašić, A. (2022).
\newblock Positioning in {5G} and {6G} networks;a survey.
\newblock {\em Sensors}, 22(13).

\bibitem[Myles et~al., 2004]{DecisionTree}
Myles, A.~J., Feudale, R.~N., Liu, Y., Woody, N.~A., and Brown, S.~D. (2004).
\newblock An introduction to decision tree modeling.
\newblock {\em Journal of Chemometrics: A Journal of the Chemometrics Society},
  18(6):275--285.

\bibitem[Noureldin et~al., 2012]{noureldin}
Noureldin, A., Karamat, T.~B., and Georgy, J. (2012).
\newblock {\em Fundamentals of inertial navigation, satellite-based positioning
  and their integration}.
\newblock Springer Science \& Business Media.

\bibitem[Refaeilzadeh et~al., 2009]{crossV}
Refaeilzadeh, P., Tang, L., and Liu, H. (2009).
\newblock Cross-validation.
\newblock {\em Encyclopedia of database systems}, 5:532--538.

\bibitem[Reid et~al., 2019]{AV}
Reid, T.~G., Houts, S.~E., Cammarata, R., Mills, G., Agarwal, S., Vora, A., and
  Pandey, G. (2019).
\newblock Localization requirements for autonomous vehicles.
\newblock {\em arXiv preprint arXiv:1906.01061}.

\bibitem[Ruan et~al., 2022]{music}
Ruan, N., Wang, H., Wen, F., and Shi, J. (2022).
\newblock Doa estimation in b{5G}/{6G}: Trends and challenges.
\newblock {\em Sensors}, 22(14).

\bibitem[Seow and Tan, 2008]{RW1}
Seow, C.~K. and Tan, S.~Y. (2008).
\newblock Non-line-of-sight localization in multipath environments.
\newblock {\em IEEE Transactions on Mobile Computing}, 7(5):647--660.

\bibitem[Wei et~al., 2011]{SB2}
Wei, X., Palleit, N., and Weber, T. (2011).
\newblock {AOD/AOA/TOA}-based {3D} positioning in {NLOS} multipath
  environments.
\newblock In {\em 2011 IEEE 22nd International Symposium on Personal, Indoor
  and Mobile Radio Communications}, pages 1289--1293.

\bibitem[Wen and Wymeersch, 2021]{SB1}
Wen, F. and Wymeersch, H. (2021).
\newblock {5G} synchronization, positioning, and mapping from diffuse
  multipath.
\newblock {\em IEEE Wireless Communications Letters}, 10(1):43--47.

\bibitem[Zhou, 2012]{Ensemble}
Zhou, Z.-H. (2012).
\newblock {\em Ensemble methods: foundations and algorithms}.
\newblock CRC press.

\end{thebibliography}
\end{document}